\documentclass[sigplan,screen,sigconf]{acmart}
\AtBeginDocument{%
  \providecommand\BibTeX{{%
    Bib\TeX}}}

\setcopyright{acmlicensed}
\copyrightyear{2025}
\acmYear{2025}

\acmConference[SC2025 (Sustainable Supercomputing Workshop)]{ACM/IEEE Conference}{November 2025}{St. Louis, MO, USA}

\usepackage{cite}
\usepackage{subcaption}
\usepackage{graphicx}
\usepackage{lipsum}
\usepackage[symbol]{footmisc}
\usepackage{amsmath,amsfonts}
\usepackage{graphicx}
\usepackage{textcomp}
\usepackage{xcolor}
\usepackage{caption}
\usepackage{enumitem}
\usepackage{adjustbox}
\usepackage{siunitx}

\usepackage{url}
\usepackage{textcomp}   
\usepackage{tikz}
\usepackage{graphicx}
\usepackage{algorithm}
\usepackage{algpseudocode}
\usepackage{caption}
\usepackage{subcaption}
\usepackage{booktabs}
\usepackage{stfloats}      
\usepackage{algorithm}
\usepackage{algpseudocode}
\def\BibTeX{{\rm B\kern-.05em{\sc i\kern-.025em b}\kern-.08em
    T\kern-.1667em\lower.7ex\hbox{E}\kern-.125emX}}

\begin{document}

\title{EMLIO: Minimizing I/O Latency and Energy Consumption \\for Large-Scale AI Training}
\author{Hasibul Jamil}

\affiliation{%
  \institution{University at Buffalo (SUNY)}
  \streetaddress{}
  \city{}
  \state{New York}
  \country{USA}
  \postcode{}
}
\email{mdhasibu@buffalo.edu}





\author{MD S Q Zulkar Nine}
\affiliation{%
  \institution{Tennessee Technological University}
  \streetaddress{}
  \city{Cookeville}
  \state{Tennessee}
  \country{USA}
  \postcode{}
}
\email{ mnine@tntech.edu}

\author{Tevfik Kosar}
\affiliation{%
  \institution{University at Buffalo (SUNY)}
  \streetaddress{}
  \city{}
  \state{New York}
  \country{USA}
  \postcode{}
}
\email{tkosar@buffalo.edu}

\renewcommand{\shortauthors}{Jamil et al.}

\begin{abstract}
Large-scale deep learning workloads increasingly suffer from I/O bottlenecks as datasets grow beyond local storage capacities and GPU compute outpaces network and disk latencies. While recent systems optimize data-loading \emph{time}, they overlook the energy cost of I/O—a critical factor at large scale. We introduce \textbf{EMLIO}, an \emph{E}fficient \emph{M}achine \emph{L}earning \emph{I/O} service that jointly minimizes end-to-end data-loading latency ($T$) and I/O energy consumption ($E$) across variable-latency networked storage. EMLIO deploys a lightweight data-serving daemon on storage nodes that serializes and batches raw samples, streams them over TCP with out-of-order prefetching, and integrates seamlessly with GPU-accelerated (NVIDIA DALI) preprocessing on the client side. In exhaustive evaluations over local disk, LAN (0.05 ms \& 10 ms RTT), and WAN (30 ms RTT) environments, EMLIO delivers up to \(\mathbf{8.6\times}\) faster I/O and \(\mathbf{10.9\times}\) lower energy use compared to state-of-the-art loaders, while maintaining constant performance and energy profiles irrespective of network distance. EMLIO's service-based architecture offers a scalable blueprint for energy-aware I/O in next-generation AI clouds.
\end{abstract}





\maketitle

\section{Introduction}

Modern AI training pipelines are increasingly data‑bound: GPUs process samples at ever higher rates, yet the data loaders -- which are responsible for fetching, decoding, and augmenting training data -- have become the performance bottleneck. Extensive work has been done on operator primitives ~\citep{Chetlur2014,Ivanov2021}, optimized computation—dedicated hardware ~\citep{Jouppi2017,Markidis2018}, high‑performance communication fabrics ~\citep{meta, hpn, Awan2017,Awan2019,Dryden2018,Nvidia2020NCCL,Sergeev2018} and compiler techniques ~\citep{Chen2018,Google2020XLA}. From the perspective of a deep‑learning framework, training a Deep Neural Network (DNN) involves three aspects: (i) computation to execute the network; (ii) communication to synchronize updates across nodes; and (iii) I/O to deliver data and labels to each worker. However, the majority of research has focused on computation and communication, shifting the bottleneck to I/O ~\citep{Murray2021,Pumma2019}. For example, when training ResNet‑50 ~\citep{resnet50} on ImageNet ~\citep{imagenet} at scale, up to 85\% of per‑epoch runtime is spent on I/O ~\citep{hdmlp2021}, and this trend holds across other workloads. As accelerators improve and datasets grow to hundreds of millions ~\citep{Sun2017} or billions ~\citep{Mahajan2018} of samples—totaling terabytes ~\citep{AbuElHaija2016,Mathuriya2018,Oyama2020} to petabytes ~\citep{Abodo2018} of storage—the I/O bottleneck will only intensify.

\begin{figure}[t]
    \centering
    \includegraphics[width=0.5\textwidth]{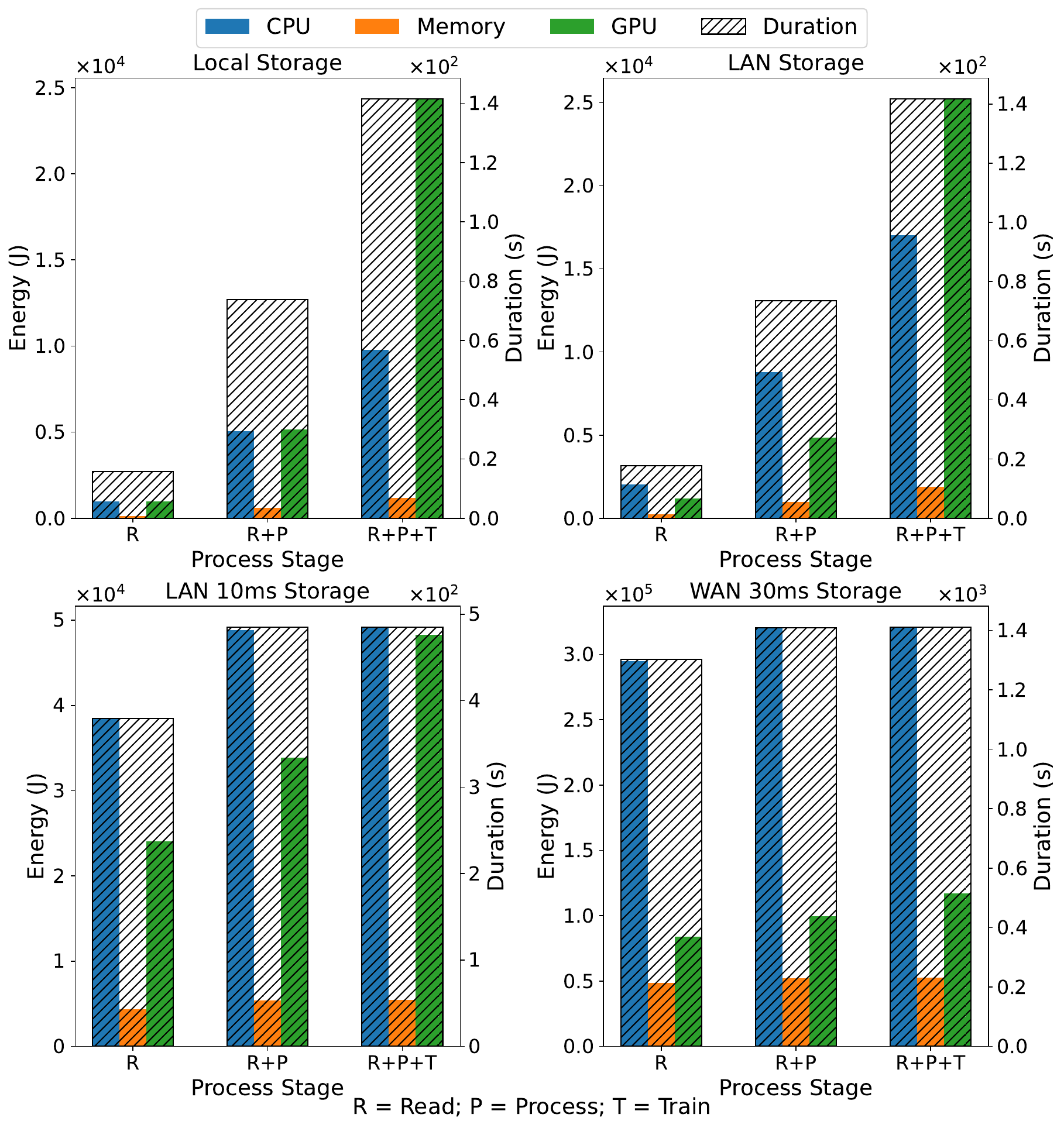}
    \caption{Breakdown of energy (CPU, DRAM, GPU) and time (hatched) for three pipeline stages (\textsc{Read}, \textsc{Read+Preprocess}, \textsc{Read+Preprocess+Train}) under four distance regimes: Local Disk, LAN (0.05 ms), emulated LAN (10 ms), and WAN (30 ms RTT). As RTT increases, I/O dominates both energy and time. All energy numbers reported also contain the ideal energy.}
  \label{fig:distance-energy-time}
\end{figure}
Optimizing training I/O is challenging because stochastic gradient descent randomly accesses small, independent samples, and shared‑filesystem contention in distributed settings further degrades performance. Existing frameworks overlap I/O with computation, but this is no longer sufficient. Ad hoc solutions—limited lookahead and double‑buffering ~\citep{Abadi2015,Chien2018,Paszke2019}, data sharding ~\citep{Goyal2017,Kurth2018}, prestaging and in‑memory caching ~\citep{oyama2020casestrongscalingdeep,Jacobs2019}, or modified access patterns ~\citep{Yang2019,Zhu2018}—suffer from limited scalability, extra hardware requirements, incomplete coverage of the storage hierarchy, or compromised dataset randomization. Indeed, GPUs often idle while waiting for data to be read and preprocessed, making I/O the true limiter of system throughput ~\citep{Mohan2021}.

Additionally, AI training at scale also incurs a tremendous energy footprint. Neural architecture search for a single NLP model can emit over 626{,}000 lbs of CO\textsubscript{2}—equivalent to five cars’ lifetime emissions—consuming hundreds of MWh of power \citep{Strubell2019}. Training GPT‑3 is estimated to require on the order of 1{,}200 MWh, yielding roughly 550 tCO\textsubscript{2}e \citep{Patterson2021}. More broadly, global data centers consumed nearly 200 TWh in 2018 $\approx$ 1\% of world electricity), with AI workloads growing fastest \citep{Koomey2020}. Crucially, data–compute locality drives both performance and energy: each additional millisecond of round‑trip time multiplies I/O energy overhead, making joint optimization of latency and energy imperative for sustainable machine learning.

Figure~\ref{fig:distance-energy-time} shows that at local storage, I/O accounts for only $\sim$15\% of energy and $\sim$20\% of per‑epoch time; however, as round‑trip latency increases, the \textsc{Read+Preprocess} stage rapidly dominates both metrics—exceeding 60\% at 10\,ms RTT and 90\% at 30\,ms RTT—highlighting that, in geo‑distributed training, data movement can dwarf compute and must be optimized for both latency and energy. These measurements were collected by training ResNet‑50~\citep{resnet50} on a 10\,GB ImageNet~\citep {imagenet} subset for one epoch, with energy and duration captured via Linux \texttt{perf} and the NVIDIA Management Library (NVML) (Section~III). Our testbed comprised three nodes provisioned on the Chameleon cloud~\citep{chameleon}: a compute node (\texttt{gpu\_rtx\_6000}) with dual Intel Xeon Gold 6126 CPUs (24 cores, 48 threads), 192 GiB DDR4, a 240 GiB SAS SSD, an NVIDIA Quadro RTX 6000 GPU, and 10 Gbps Ethernet; a storage node (\texttt{compute\_skylake}) with identical CPU, memory, storage, and networking but without a GPU; and another storage node (\texttt{storage}) with dual Intel Xeon E5‑2650 v3 CPUs (20 cores, 40 threads), 64 GiB DDR4, a 400 GiB SATA SSD, and 10 Gbps Ethernet.

Several systems advance I/O throughput—NVIDIA DALI ~\citep{NVIDIA2018DALI} and FFCV~\citep{ffcv} for local storage, and network‑aware services such as NoPFS \citep{hdmlp2021}, Lobster \citep{lobster}, and Cassandra-DALI ~\citep{nosqlItaly} for remote datasets—but none quantify or minimize the energy cost of data movement, decoding, and preprocessing. Ignoring energy consumption overlooks a substantial portion of the total training footprint, particularly in geo‑distributed scenarios.

In this paper, we present \textbf{EMLIO}—\emph{E}fficient \emph{M}achine \emph{L}earning \emph{I/O}—a service‑based framework that jointly minimizes data‑loading latency and I/O energy consumption. EMLIO co‑locates a lightweight daemon with storage servers to asynchronously fetch raw samples, serialize them into pre‑batched payloads, and stream them over TCP over the network fabric to the compute nodes. Its out‑of‑order prefetching pipeline exploits network and storage parallelism to bound tail latency under high RTT. To our knowledge, this is the first work to quantify end‑to‑end I/O energy consumption during AI model training and to optimize both performance and energy without sacrificing throughput.

The major contributions of EMLIO include:
\begin{itemize}[leftmargin=*]
  \item \emph{Service‑based I/O offload}, decoupling data movement from framework internals via a lightweight storage daemon.
  \item \emph{Out‑of‑order prefetching}, bounding I/O tail latency across RTTs by exploiting parallelism.
  \item \emph{Energy‑aware evaluation}, the first end‑to‑end measurement and optimization of I/O energy in ML, achieving up to 10.9× reduction.
  \item \emph{Performance retention}, demonstrating state‑of‑the‑art epoch times across local, LAN, and WAN deployments.
\end{itemize}

The remainder of the paper is organized as follows. Section II reviews related work. Section III describes our distributed energy measurement framework. Section IV details the design and implementation of EMLIO. Section V evaluates its performance and energy efficiency across diverse storage fabrics. Finally, Section VI concludes and outlines directions for future work.

\section{Related Work}

\paragraph{Local‐ and Mounted Remote‐Storage Optimizations}  
FFCV~\citep{ffcv} and NVIDIA DALI~\citep{NVIDIA2018DALI} accelerate I/O on SSD‐backed file systems (including remote repositories exposed via file‐system mounts) by combining custom data formats with offloaded transformations. FFCV’s memory‐mapped “.beton” format and JIT‐compiled preprocessing eliminate host‐side load/store overhead, while DALI performs image decoding and augmentation directly on the GPU. Both frameworks report up to 30× reduction in epoch time on local disks; however, they do not quantify I/O energy and exhibit degraded performance when accessing non‑local datasets (see Section V).

\paragraph{Network‐Aware Prefetching}  
Clairvoyant Prefetching (NoPFS)~\citep{hdmlp2021}, Lobster~\citep{lobster}, and extend data loaders to remote storage. Cassandra‐DALI~\citep{nosqlItaly} masks network latency: NoPFS employs out‐of‐order block prefetching, Cassandra‐DALI leverages NoSQL backends to hide RTT, and Lobster uses cooperative caching with adaptive prefetch windows to sustain high WAN throughput. These techniques can reduce end‑to‑end latency by up to 5× in geo‑distributed training, but they do not account for the energy consumed during I/O.  

\paragraph{Energy‐Aware I/O}  
Outside of machine learning, several systems optimize energy in storage and networking. GreenTCP~\citep{Wang2018GreenTCP} dynamically tunes TCP congestion windows to trade throughput for NIC energy savings, and EcoCache~\citep{Patel2019EcoCache} schedules disk spin‑down to reduce HDD power in distributed filesystems. Jamil et al.~\citep{jamil2025} introduce a dynamic, multiparameter reinforcement‑learning framework that adaptively adjusts application‑layer transfer settings—pausing and resuming threads based on network load—to balance high throughput, fairness, and energy efficiency, achieving up to 25\% higher throughput and 40\% lower energy use. However, these approaches target general data services and do not meet the low‑latency, high‑throughput, energy‑critical demands of AI training I/O.

Unlike prior ML data loaders that focus solely on throughput and latency—FFCV~\citep{ffcv} and NVIDIA DALI~\citep{NVIDIA2018DALI} for local disks, or network‑aware loaders like NoPFS~\citep{hdmlp2021} and Lobster~\citep{lobster}, Cassandra-DALI~\citep{nosqlItaly} for remote datasets—EMLIO targets a combined time–energy objective by explicitly measuring and minimizing I/O energy across storage and compute resources while preserving low per‑epoch latency. EMLIO’s architecture rests on three complementary techniques: (i) {TFRecord sharding}, which stores data in large TFRecord files and assembles training batches by randomly sampling within each shard—eliminating the overhead of small‐file reads and excessive metadata operations; (ii) {service‐based batching with out‐of‐order prefetching}, where a storage‐side daemon aggregates and serializes entire batches before streaming over TCP and leverages parallel prefetch pipelines to fully utilize network bandwidth; and (iii) {RTT‐resilient adaptivity},  maintain near‐constant I/O time and energy profiles (±5\%) from sub‑millisecond LANs to 30 ms WANs, in contrast to existing loaders whose efficiency degrades under high latency. Together, these advances make EMLIO a necessary step toward sustainable, scalable AI training in geo‑distributed and energy‑constrained environments.


\section{Distributed Energy Measurement Framework}

To quantify and optimize energy consumption at system‐level granularity, we developed \texttt{EnergyMonitor}, a fine‐grained measurement framework that captures both component‐level (CPU package, DRAM, GPU) ideal energy and process‐specific usage energy across geo‐distributed nodes. Each node runs synchronized sampling threads for CPU/DRAM and GPU power, taking snapshots every 100 ms—fast enough to capture dynamic AI workload behavior ~\citep{fecom}—while using a threading barrier to align all samplers on the same timestamp \(t_k\). This exact time alignment ensures that, for each \(t_k\), we obtain a coherent energy tuple spanning all components. 

If a sampler misses its interval (e.g., due to transient delay), \\
\texttt{EnergyMonitor} automatically interpolates the missing values, preventing gaps in the time series. Because all nodes synchronize their clocks via NTP, we can later query our time‐series database TSDB (e.g., InfluxDB) for any known start and end timestamps and accurately aggregate each node’s energy consumption over that interval.

\begin{figure}[t]
  \centering
  \includegraphics[width=\columnwidth]{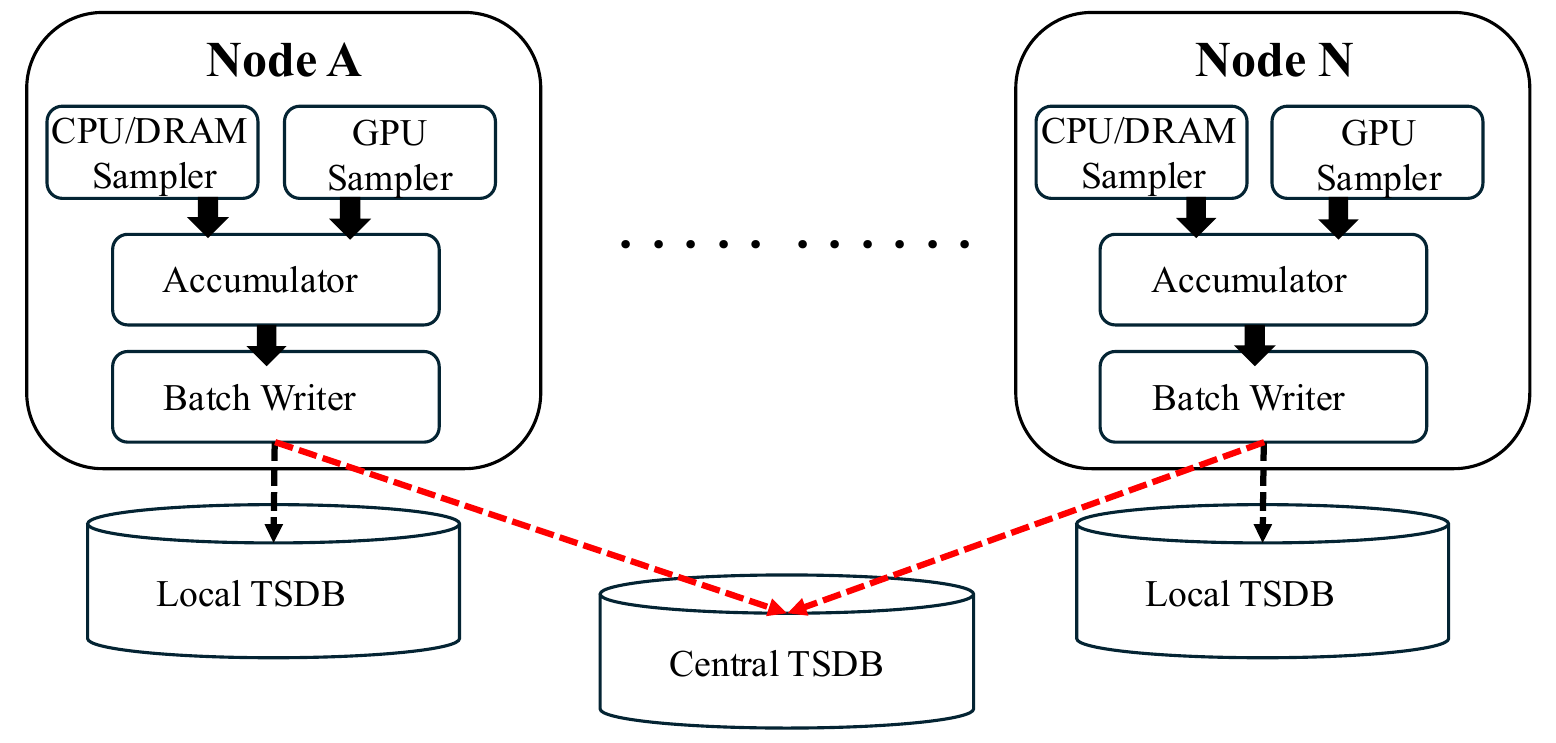}
  \caption{Block diagram of the distributed energy‐monitoring framework. Each Compute Node runs two synchronized samplers (CPU/DRAM and GPU), an Accumulator, and a Batch Writer. Local writes go to a per‐node time series database (TSDB), or the data could be forwarded to a central TSDB. All system clocks are kept in sync via NTP.}
  \label{fig:energy-arch}
\end{figure}

Figure~\ref{fig:energy-arch} shows the overall architecture: per‐node CPU/DRAM and GPU samplers synchronize on a barrier, enqueue timestamped energy readings, and forward them to an Accumulator that merges and interpolates by \(t_k\). A Batch Writer then periodically writes these aligned tuples—tagged by node ID—to the TSDB for cross‐node analysis.  

\begin{algorithm}[!t]
\scriptsize 
\caption{Distributed Energy Sampling and Reporting}
\label{alg:energy}
\begin{algorithmic}[1]
 \Require Sampling interval $\delta$, node id $\mathit{id}$, InfluxDB config 
 \Ensure Time‑aligned energy tuples in InfluxDB tagged by $\mathit{id}$
 \State \textbf{Initialize:}
 \Statex\quad 1. \texttt{nvmlInit()}; detect GPUs and handles.
 \Statex\quad 2. Connect to InfluxDB.
 \Statex\quad 3. Create a \texttt{Barrier} for $1+[\,\mathit{hasGPU}\,]$ threads.
 \State \textbf{Launch Threads:} CPU/DRAM sampler, optional GPU sampler, accumulator, writer.
\While{monitoring active}
\State \textbf{Sampler threads:}
\Statex\quad Barrier‑wait to align \(t_k\).
\If{CPU/DRAM}
\State run \texttt{perf stat -e power/energy-pkg/,power/energy-ram/ sleep $\delta$}
\State parse $(E_{\mathrm{pkg}},E_{\mathrm{ram}})$
\State enqueue $(t_k,\{\texttt{cpu\_energy}=E_{\mathrm{pkg}},\,\texttt{memory\_energy}=E_{\mathrm{ram}}\})$
\EndIf
\If{GPU}
\State read each $P_i$ via NVML, compute $E_{\mathrm{gpu}}=\sum_i P_i\delta/1000$
\State enqueue $(t_k,\{\texttt{gpu\_energy}=E_{\mathrm{gpu}}\})$
\EndIf
\State \textbf{Accumulator:}
\Statex\quad merge CPU/DRAM+GPU by $t_k$, interpolate holes, forward tuples
\State \textbf{Writer:}
\Statex\quad batch up to $N$ tuples, tag with \texttt{node\_id}, call \texttt{write\_points()}
\EndWhile
\State shutdown threads and NVML cleanly
\end{algorithmic}
\end{algorithm}

\subsection*{Explanation of Algorithm~\ref{alg:energy}}

\paragraph{Lines 1 (Initialization)} 
Initialize NVML, TMDS client, and a threading barrier (see Figure~\ref{fig:energy-arch}) so that CPU/DRAM and GPU samplers share a common timestamp $t_k$.

\paragraph{Lines 2 (Thread Launch)} 
As shown in Figure~\ref{fig:energy-arch}, each node starts:
\begin{itemize}
\item \emph{CPU/DRAM sampler} and (optionally) \emph{GPU sampler} 
\item \emph{Accumulator} that merges per‐component queues 
\item \emph{Batch Writer} that writes locally or forwards to the central TSDB 
\end{itemize}

\paragraph{Lines 3–13 (Sampling Loop)} 
The two samplers synchronize on the barrier (Figure~\ref{fig:energy-arch}) to align at $t_k$, measure via \texttt{perf stat} or NVML, enqueue their tuples, then wake the Accumulator.

\paragraph{Lines 14-16 (Accumulator \& Writer)} 
The Accumulator (middle box in Figure~\ref{fig:energy-arch}) merges and fills missing slots; the Batch Writer (bottom box) writes batches of up to $N$ points either to the local TSDB or into the central TSDB for cross‐node correlation.

\paragraph{Line 17 (Shutdown)} 
On termination, all threads are joined, and NVML is cleanly shut down.

\subsection*{Implementation Notes}

This design depends on three properties:
\begin{itemize}
 \item {100 ms sampling interval:} sufficient resolution for CPU package, DRAM, and GPU power dynamics. 
 \item {Barrier‐synchronized threads:} guarantee component‐aligned readings at each \(t_k\). 
 \item {Interpolation:} fills any missed samples to maintain a gapless time series. 
 \item {Time‐synchronized nodes:} NTP alignment allows event‐level queries over distributed workloads by specifying start/end timestamps in the TSDB. 
\end{itemize}
We utilized standard interfaces—Linux \texttt{perf stat} for CPU/DRAM ~\citep{KernelPerfAdminGuide}, NVIDIA NVML API v5.0 for GPU ~\citep{NvidiaNVML}, and InfluxDB v1.8 Python client ~\citep{InfluxDBPythonDocs} —to ensure portability and maintainability across diverse cluster environments.

\section{EMLIO Design Principles and Implementation}

EMLIO is built around three core design principles:

\begin{enumerate}
  \item {Network–pipeline concurrency.}  All I/O stages—reading, serialization, and network send—run in parallel on separate threads, so that while one batch is being sent over the network, the next batch is already being read and prepared, fully utilizing available bandwidth and hiding per‑batch latency.
  \item {Batch‑aligned data‑parallel planning.}  A centralized \emph{Planner} computes, for each epoch and compute node, exactly which TFRecord \citep{TensorFlowTFRecord} shard ranges form each fixed‑size batch.  This ensures correct data‑parallel semantics without client‑side shard scans or random small reads.
  \item {Linear horizontal scalability.} Each storage node runs its own EMLIO Daemon with a pool of worker threads.  By sharding TFRecords and parallelizing both read/serialize and send pipelines, EMLIO’s aggregate throughput grows linearly as storage nodes are added.
\end{enumerate}

\subsection{High-Level Architecture}

\begin{figure}[t]
  \centering
  \includegraphics[width=\columnwidth]{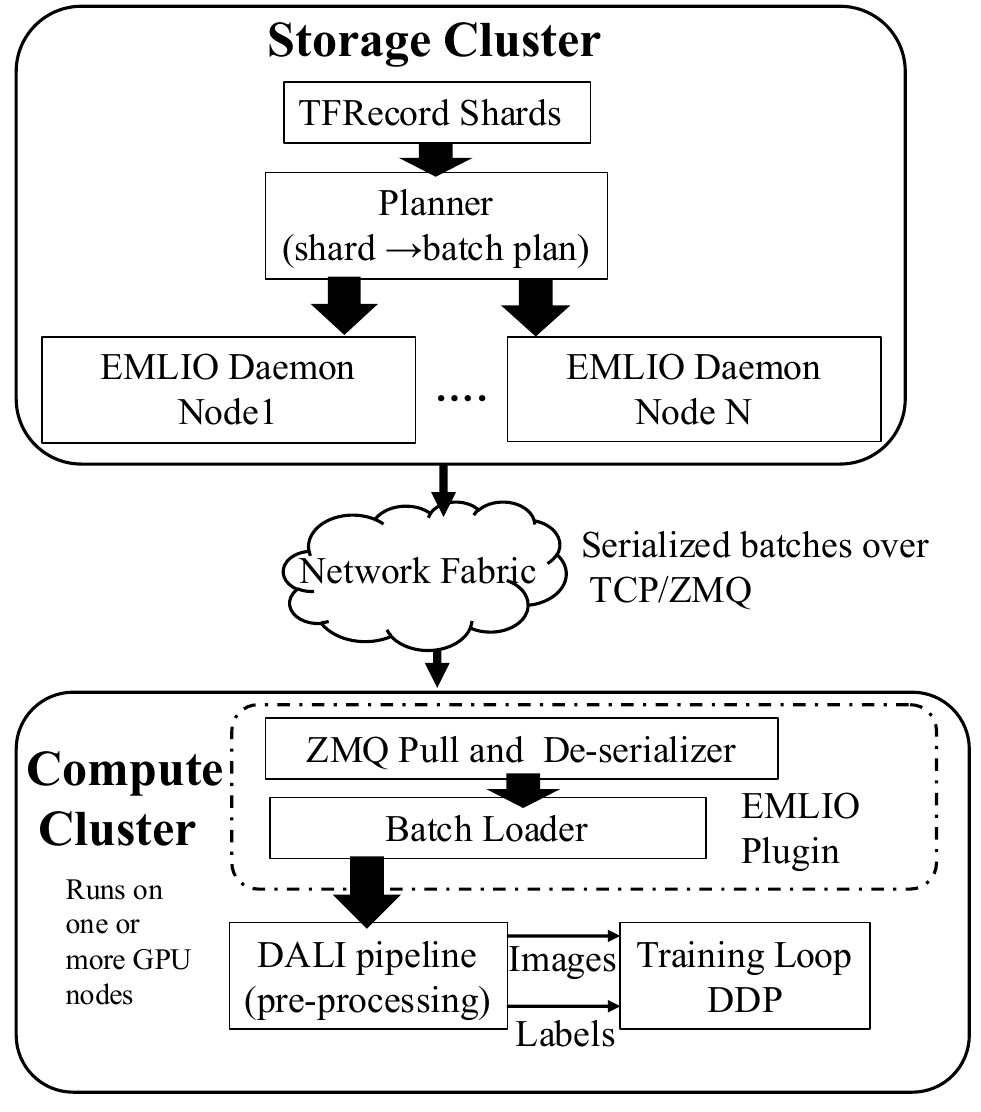}
  \caption{Block diagram of the EMLIO sender (storage side) and receiver (compute side).}
  \label{fig:emlio-arch}
\end{figure}

The overall architecture of EMLIO, illustrated in Figure~\ref{fig:emlio-arch}, consists of two major components: the \emph{storage side} (sender) and the \emph{compute side} (receiver), connected via a high-performance network fabric.

\paragraph{Storage side.}
A global \emph{Planner} ingests TFRecord shard metadata (including file paths and byte offsets), the list of available compute nodes (IP addresses and ports), and epoch/batch-size parameters. Based on these inputs, it produces a \emph{batch plan} mapping contiguous TFRecord offsets into batches for each node and epoch. Each storage server runs an \emph{EMLIO Daemon} that reads its assigned TFRecord shards via \texttt{mmap}, serializes groups of $B$ examples into a single \texttt{msgpack} payload, and pushes these pre-batched payloads over ZeroMQ streams according to the plan. Here, \texttt{msgpack} is a compact, binary serialization format that is both fast and space-efficient, enabling high-throughput encoding and decoding of structured data. By decoupling planning from the I/O daemons and multi-threading both the read/serialize and send stages, EMLIO fully pipelines disk and network operations, maximizing end-to-end throughput even under high round-trip times.




\paragraph{Network fabric.}  
The serialized Batches traverse multiple TCP/ZeroMQ streams with built‑in backpressure (high‑water marks), automatically adapting to receiver speed and smoothing out per‑batch latency fluctuations.

\paragraph{Compute side.}
Each GPU node runs an \emph{EMLIO Receiver} that pulls incoming msgpack batches into a shared in-memory queue. \\
A \texttt{BatchProvider} deserializes each payload and exposes the samples as DALI’s \texttt{external\_source}. The DALI pipeline then performs GPU-accelerated preprocessing—decoding JPEGs, resizing, cropping, normalizing tensors, and asynchronously prefetching multiple batches—before handing the data to the PyTorch Distributed Data Parallel (DDP) training loop. Because compute nodes work entirely on large, in-memory pre-batched data rather than issuing small, random reads, EMLIO achieves both higher throughput and significantly reduced I/O-related energy consumption per epoch.

\subsection{Planning and Dispatch Algorithm}

\begin{algorithm}[h]
\scriptsize
\caption{EMLIO Sender Planning \& Dispatch}
\label{alg:emlio-sender}
\begin{algorithmic}[1]
  \Require TFRecord directory $D$, node list $N=\{(id_i,ip_i,port_i)\}$, 
           batch size $B$, epochs $E$, threads per node $T$
  \Ensure Each compute node receives exactly $E\times\lceil |D|/B\rceil$ batches
  \State \textbf{Load shard metadata:}
    \Statex\quad parse \texttt{mapping\_shard\_*.json} to get offsets/sizes
  \State \textbf{Build global label map} from all shards
  \For{$e=1$ \textbf{to} $E$}
    \State shuffle shard list
    \State assign shards to nodes round‐robin
    \For{each node $(id,ip,port)$}
      \State split its shard list into $T$ subsets
      \State launch $T$ threads (ThreadPoolExecutor) with \textproc{SendWorker}
    \EndFor
  \EndFor
\end{algorithmic}
\end{algorithm}

\paragraph{Component Explanation}
\begin{itemize}
  \item {Lines 1–2:} Read TFRecord shard index files to build a global map of $(\mathrm{offset},\mathrm{size},\mathrm{label})$ tuples.  
  \item {Lines 3–4:} Shuffle for each epoch to ensure randomness; assign shards evenly to nodes to satisfy data‐parallel epoch coverage.  
  \item {Lines 5–8:} Each node’s work is further subdivided into $T$ threads.  \textproc{SendWorker} (see code) mmaps its assigned TFRecord, batches $B$ records by slicing and msgpack‐serializing, and PUSHes them over ZeroMQ—implicitly providing backpressure via ZMQ HWM.  
\end{itemize}

\subsection{ Rationale For Using TFRecord Format}

We chose the TensorFlow TFRecord format because it makes loading batches both simple and efficient:

\begin{itemize}[leftmargin=*]
  \item {Fast, contiguous reads:} TFRecord files store examples one after another, each prefixed by its length. By memory‐mapping the file, we can grab a block of $B$ examples in one go—no extra copies or individual read calls.
  \item {Lower I/O overhead:} Reading large, sequential chunks instead of many small, random reads cuts down on kernel and filesystem work, speeding up data loading and reducing latency.
\end{itemize}

The upfront cost of converting raw data into TFRecord format is incurred only once and is amortized across all subsequent training jobs that use the same dataset.

\subsection{Receiver Algorithm}

\begin{algorithm}[h]
\scriptsize
\caption{EMLIO Receiver \& DALI Integration}
\label{alg:emlio-receiver}
\begin{algorithmic}[1]
  \Require bind address \texttt{ip}, port \texttt{p}, prefetch depth $Q$, epochs $E$
  \Ensure In‐GPU ready batches fed into DDP training
  \State start ZMQ PULL socket on (\texttt{ip},\texttt{p})
  \State spawn \texttt{zmq\_receiver} thread to push into shared Queue
  \State build DALI pipeline with \texttt{BatchProvider}(Queue), prefetch=$Q$
  \State \textbf{warm up:} manually run $Q$ iterations
  \For{epoch = 1 \textbf{to} $E$}
    \While{data is available}
      \State \texttt{pipe.run()} \Comment{deserializes \& preprocesses}
      \State feed output to DDP training step
    \EndWhile
  \EndFor
  \State teardown pipelines and sockets
\end{algorithmic}
\end{algorithm}

\paragraph{Component Explanation}
\begin{itemize}
  \item {Lines 1–2:} Bind a ZeroMQ PULL socket, spawn a thread that unpacks msgpack batches into a shared Queue.  
  \item {Lines 3–4:} Define a DALI pipeline whose \texttt{external\_source} pulls from the Queue in parallel.  DALI’s GPU‐accelerated decoding and normalization then proceed in the pipeline.  
  \item {Lines 5–9:} Warm up the pipeline to fill internal buffers, then iterate epoch‐wise, running the pipeline and passing its output to the training loop (DistributedDataParallel, optimizer, loss, backward, step).  
\end{itemize}

\subsection{Implementation Details}

\paragraph{ZeroMQ for Backpressure}  
We set the PUSH socket’s HWM to 16 and blocking send to infinity, ensuring that storage‐side workers naturally back off when compute‐side queues are full.

\paragraph{Concurrency Model}  
On the sender side, each TFRecord shard is processed by up to $T$ threads in parallel.  On the receiver, DALI’s $exec\_async$ and $exec\_pipelined$ options run decode/Augment and CPU preprocessing concurrently with GPU execution.

\paragraph{Timestamp Logging}  
Both sender and receiver log events (batch send, batch receipt, epoch start/end) via a shared $TimestampLogger$ utility, enabling post‐hoc alignment with the energy‐monitoring traces in InfluxDB.


\section{Evaluation}

We evaluate EMLIO under three scenarios:
\begin{enumerate}[leftmargin=*]
  \item {Centralized repository:} All training data is hosted on a single remote NFSv4 server.
  \item {Fully sharded:} Data is evenly distributed across all participant nodes, eliminating any central storage point.
  \item {Outcome comparison:} We measure the impact of I/O on training accuracy, convergence speed, and energy efficiency.
\end{enumerate}

\subsection{Scenario 1: Centralized Repository}
\label{sec:eval-scenario1}

\subsubsection{Setup}
\begin{figure}[t]
  \centering
  \includegraphics[width=0.8\columnwidth]{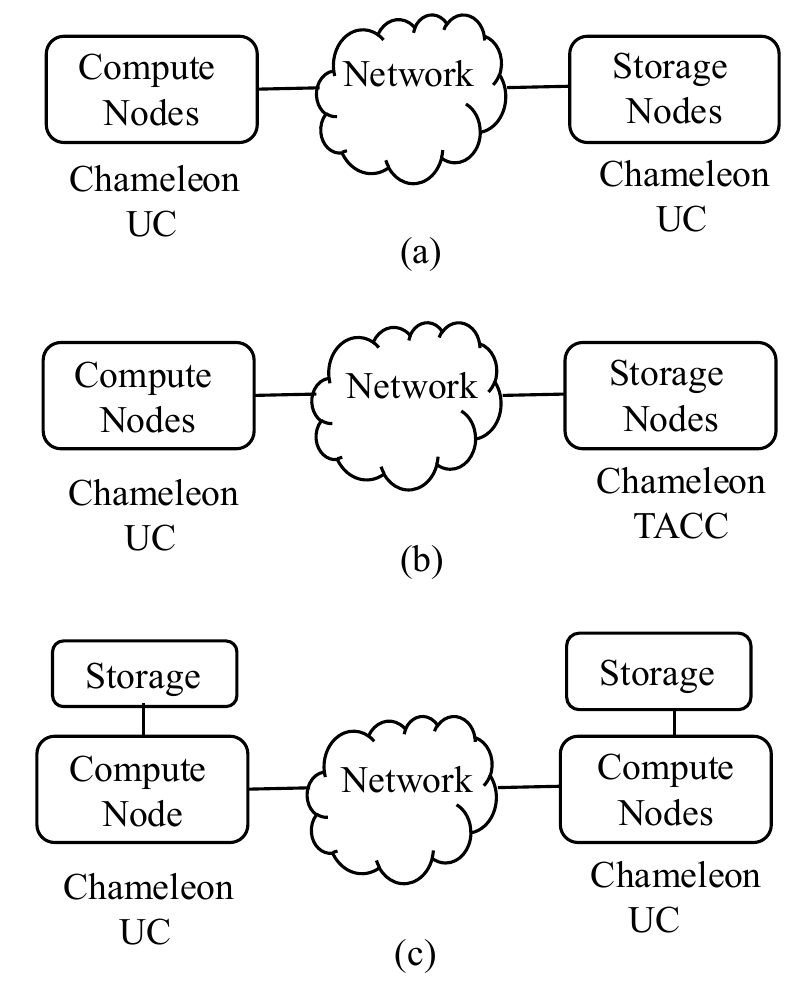}
  \caption{Topologies for our three evaluation scenarios:
    (a) centralized storage in the UC Chameleon cloud (0.1 ms RTT),
    (b) cross‑site WAN between UC compute and TACC storage (30 ms RTT),
    and (c) fully sharded data across compute nodes.}
  \label{fig:evaluation_setup}
\end{figure}

In this scenario, all training data (ImageNet~\citep{imagenet}, COCO~\citep{coco}, or synthetic) reside on a single NFSv4‑mounted storage server. We test four network regimes:  
\begin{itemize}[noitemsep,leftmargin=*]
  \item {Local storage} (data on local disk),  
  \item {LAN (0.1\,ms RTT)},  
  \item {Emulated LAN} (1\,ms RTT , 10\,ms RTT, 30\,ms RTT injected via Linux \texttt{tc}/\texttt{qdisc}),  
  \item {WAN (30\,ms RTT)}.
\end{itemize}
Compute nodes always access data over NFSv4. Figure~\ref{fig:evaluation_setup}(a--b) illustrates the LAN and WAN topologies used in our experiments. Refer to Table~\ref{tab:node-specs} for detailed specifications of the compute and storage nodes used in the evaluation.

We compare three data‑loading pipelines:
\begin{itemize}[noitemsep,leftmargin=*]
  \item {PyTorch DataLoader} reading directly over NFSv4;
  \item {NVIDIA DALI} pipeline over NFSv4;
  \item {EMLIO}, with EMLIO Daemon on the storage node \\ 
  (mmap→serialize→ZMQ push) and EMLIO Receiver into a DALI pipeline on the compute side.
\end{itemize}

Each pipeline trains ResNet‑50 ~\citep{resnet50} for one epoch on three workloads:
\begin{enumerate}[noitemsep,leftmargin=*]
  \item \emph{ImageNet} (0.1 MB/sample),
  \item \emph{COCO} (0.2 MB/sample),
  \item \emph{Synthetic} (2 MB/sample).
\end{enumerate}
We measure per‑epoch duration, CPU/DRAM/GPU energy (via our distributed monitor), and overall throughput.

\begin{figure}[t]
  \centering
  \includegraphics[width=\columnwidth]{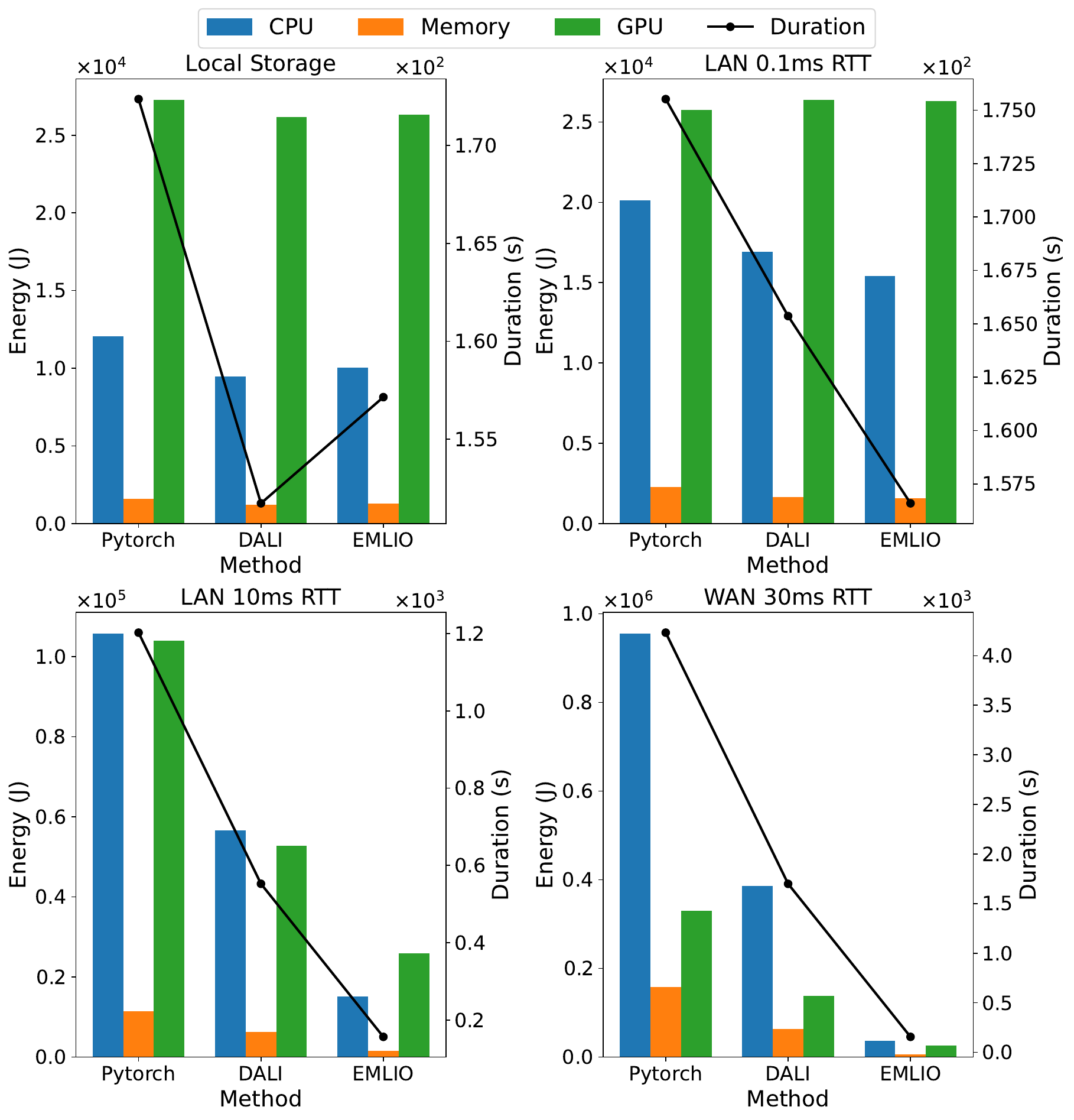}
  \caption{
    Per‑epoch energy (CPU, DRAM, GPU) and runtime for PyTorch DataLoader, NVIDIA DALI, and EMLIO across four regimes: local disk, LAN (0.1 ms), LAN (10 ms), and WAN (30ms) for the ImageNet dataset. EMLIO maintains near‑constant duration and minimal I/O energy as RTT increases, while both PyTorch and DALI incur steep penalties at higher latencies.
  }
  \label{fig:eval-imgnet}
\end{figure}
\begin{table}[t]
  \centering
  \caption{Node specifications for UC and TACC clusters in chameleon ~\citep{chameleon} cloud. All nodes run Ubuntu 22.04 LTS, CUDA 12.6 (only in GPU nodes), and mount training data via NFSv4.}
  \label{tab:node-specs}
  \scriptsize
  \begin{adjustbox}{max width=\columnwidth}
  \begin{tabular}{@{}ll@{}}
    \toprule
    \textbf{Node Type} & \textbf{Specification} \\
    \midrule
    UC Compute\\(\texttt{gpu\_rtx\_6000}) &
      \begin{minipage}[t]{0.8\columnwidth}
      \begin{itemize}[leftmargin=*]
        \item \textbf{CPU:} 2× Intel Xeon Gold 6126 @2.60 GHz (12 cores/socket, 24 HT/socket; 48 threads)
        \item \textbf{RAM:} 192 GiB DDR4
        \item \textbf{Storage:} 240 GiB SAS SSD (MZ7KM240HMHQ0D3)
        \item \textbf{Network:} 10 Gbps Ethernet (Intel X710 SFP+, 2 ports)
        \item \textbf{GPU:} 1× NVIDIA Quadro RTX 6000 (TU102GL, 24 GiB)
      \end{itemize}
      \end{minipage}
      \\[\abovecaptionskip]
    UC Storage\\(\texttt{compute\_skylake}) &
      \begin{minipage}[t]{0.8\columnwidth}
      \begin{itemize}[leftmargin=*]
        \item \textbf{CPU:} 2× Intel Xeon Gold 6126 @2.60 GHz (12 cores/socket, 48 threads)
        \item \textbf{RAM:} 192GiB DDR4
        \item \textbf{Storage:} 240 GiB SAS SSD (MZ7KM240HMHQ0D3)
        \item \textbf{Network:} 10 Gbps Ethernet (Intel X710 SFP+, 2 ports)
        \item \textbf{GPU:} None
      \end{itemize}
      \end{minipage}
      \\[\abovecaptionskip]
    TACC Compute\\(\texttt{gpu\_p100}) &
      \begin{minipage}[t]{0.8\columnwidth}
      \begin{itemize}[leftmargin=*]
        \item \textbf{CPU:} 2× Intel Xeon E5‑2670 v3 @2.30 GHz (12 cores/socket, 48 threads)
        \item \textbf{RAM:} 128 GiB DDR4
        \item \textbf{Storage:} 1 TB SATA HDD (ST1000NX0443)
        \item \textbf{Network:} 10 Gbps Ethernet (Broadcom NetXtreme II BCM57800, 1 port)
        \item \textbf{GPU:} 2× NVIDIA Tesla P100 (16 GiB each)
      \end{itemize}
      \end{minipage}
      \\[\abovecaptionskip]
    TACC Storage\\(\texttt{storage}) &
      \begin{minipage}[t]{0.8\columnwidth}
      \begin{itemize}[leftmargin=*]
        \item \textbf{CPU:} 2× Intel Xeon E5‑2650 v3 @2.30 GHz (10 cores/socket, 40 threads)
        \item \textbf{RAM:} 64 GiB DDR4
        \item \textbf{Storage:} 400 GiB SATA SSD (INTEL SSDSC1BG40)
        \item \textbf{Network:} 10 Gbps Ethernet (Broadcom NetXtreme II BCM57810, 1 port)
        \item \textbf{GPU:} None
      \end{itemize}
      \end{minipage}
      \\
    \bottomrule
  \end{tabular}
  \end{adjustbox}
\end{table}

\paragraph{ImageNet 10\,GB Subset}  
Figure~\ref{fig:eval-imgnet} plots per‑epoch energy (bars: CPU, DRAM, GPU) and runtime (line) for ResNet‑50 training on a 10 GB ImageNet subset under four network regimes: local disk, LAN (0.1 ms RTT), LAN (10 ms RTT), and WAN (30 ms RTT). Key observations:

\begin{itemize}[leftmargin=*]
  \item {Local disk:} DALI leads at 151.7s, PyTorch trails at 172.4s, and EMLIO finishes in 157.1s (2\% slower than DALI). This slight overhead arises because EMLIO still pipelines through the network stack even when storage and compute are on the same node: it reads from disk, sends data through the loopback network interface, and re‑ingests it, adding CPU load. CPU energy ranges from 9.5–12.0 kJ, GPU energy is 26.2–27.3 kJ, and DRAM energy remains below 1.3 kJ for all methods.
  \item {LAN (0.1 ms):} EMLIO holds at 156.6 s, while DALI and PyTorch rise to 165.4 s and 175.5s. EMLIO’s CPU ( 10.1 kJ) and GPU ( 26.3kJ) energies remain close to local values, whereas PyTorch and DALI see modest increases.
  \item {LAN (10ms):} EMLIO finishes in 156.5s, but DALI and PyTorch slow dramatically to 552.5s and 1202.2s. Correspondingly, EMLIO’s CPU (9.9 kJ) and GPU (25.9 kJ) remain steady, while DALI’s and PyTorch’s energies climb by 3×–4×.
  \item {WAN (30 ms):} EMLIO still completes in 156.2 s, whereas DALI and PyTorch require 1699.3 s and 4232.4 s. EMLIO’s energy stays at ~10.0kJ (CPU) and ~26.2kJ (GPU), but DALI’s and PyTorch’s energies increase by up to 60× under high latency.
\end{itemize}

These results demonstrate that EMLIO’s pipelined batching and multi‑stream push fully mask network delays: its epoch time varies less than 5\% and its I/O‑related energy remains minimal from 0.1 ms to 30 ms RTT, whereas both DALI and PyTorch incur 3×–27× longer runtimes and 4×–60× higher energy as latency increases.

\begin{figure}
  \centering
  \includegraphics[width=\columnwidth]{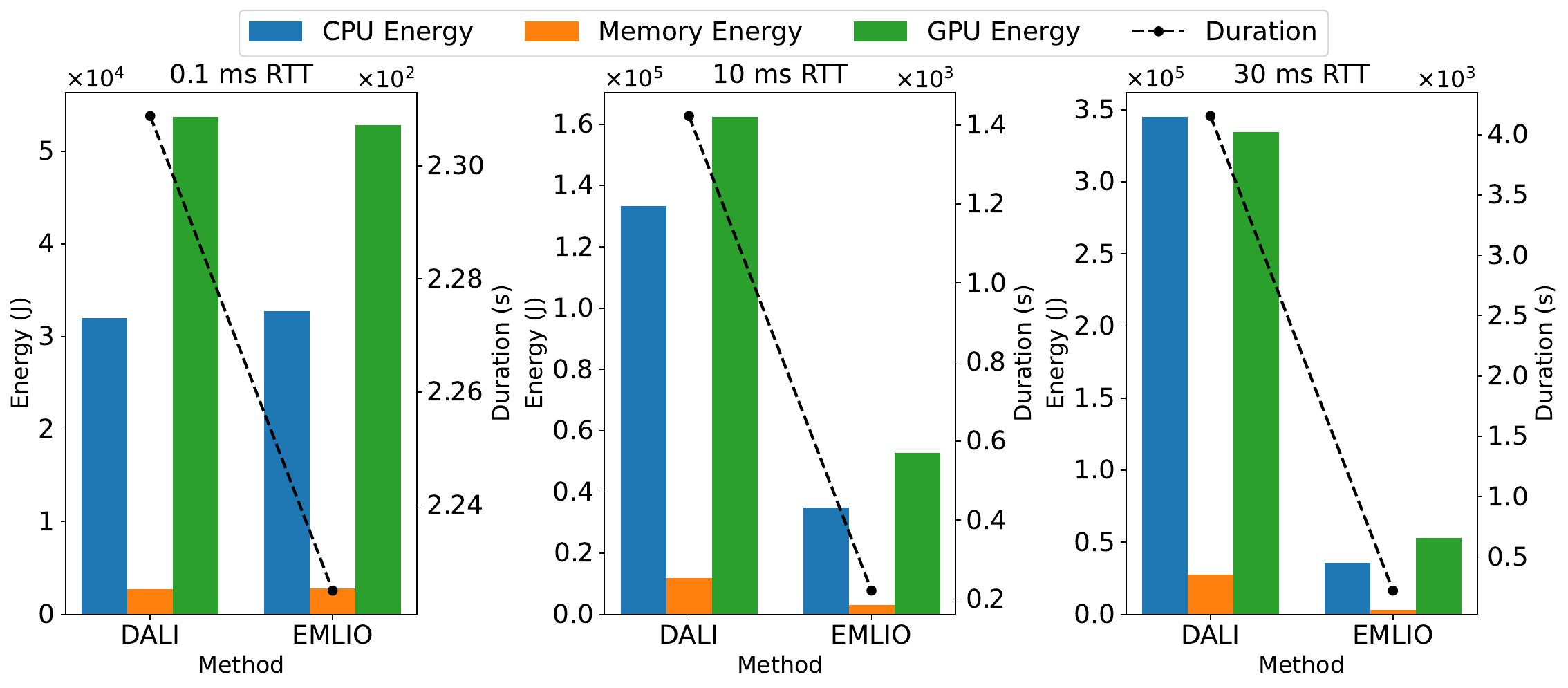}
   \caption{
    Breakdown of per‐epoch energy consumption (CPU, DRAM, GPU) and runtime (line) 
    for NVIDIA DALI, and EMLIO across three network RTT regimes:
    LAN at 0.1 ms, LAN at 10 ms, and LAN at 30 ms for the COCO dataset.
    EMLIO maintains nearly constant duration and minimal I/O energy usage as RTT increases,
    whereas DALI suffers steep increases in time and energy at higher latencies.
  }
  \label{fig:eval-coco}
\end{figure}

\paragraph{COCO}  
Since PyTorch DataLoader performed poorly (see Figure~\ref{fig:eval-imgnet}), we restrict comparison to EMLIO and DALI. The COCO workload (Figure~\ref{fig:eval-coco}) follows the same trend: EMLIO maintains stable throughput and nearly constant I/O energy from LAN through WAN, whereas DALI’s performance and energy efficiency both degrade steadily as RTT increases. Because COCO samples are slightly smaller, the relative benefit grows: at 30ms RTT, EMLIO is roughly 6× faster and consumes 8× less I/O energy than DALI.

\begin{figure}
  \centering
  \includegraphics[width=\columnwidth]{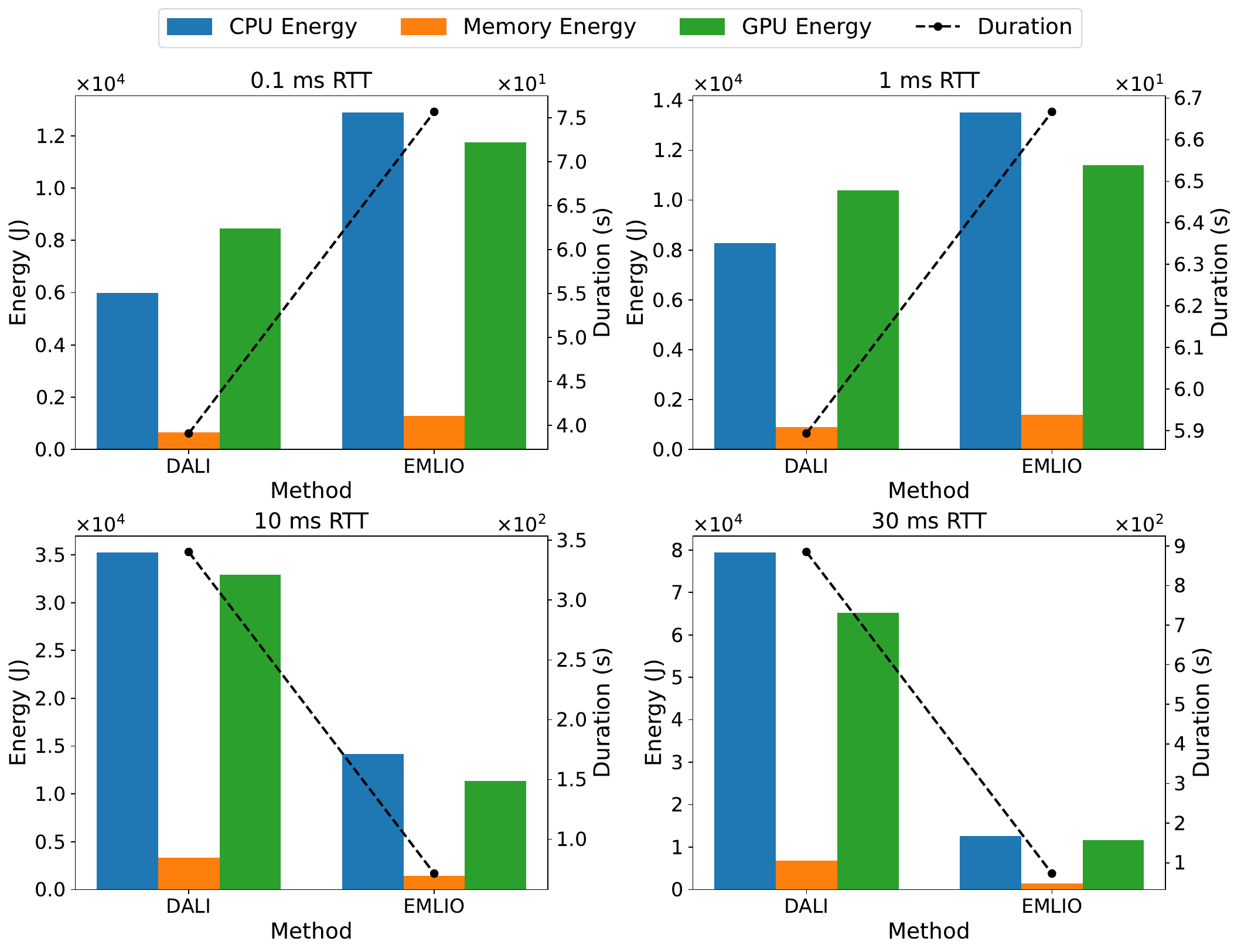}
  \caption{
    Breakdown of per‐epoch energy consumption (CPU, DRAM, GPU) and runtime (line) 
    for NVIDIA DALI, and EMLIO across three network RTT regimes:
    LAN at 0.1 ms, LAN at 1 ms, LAN at 10 ms, and LAN at 30 ms for the synthetic dataset. EMLIO employs concurrency 1. 
  }
  \label{fig:eval-synth}
\end{figure}

\begin{figure}
  \centering
  \includegraphics[width=\columnwidth]{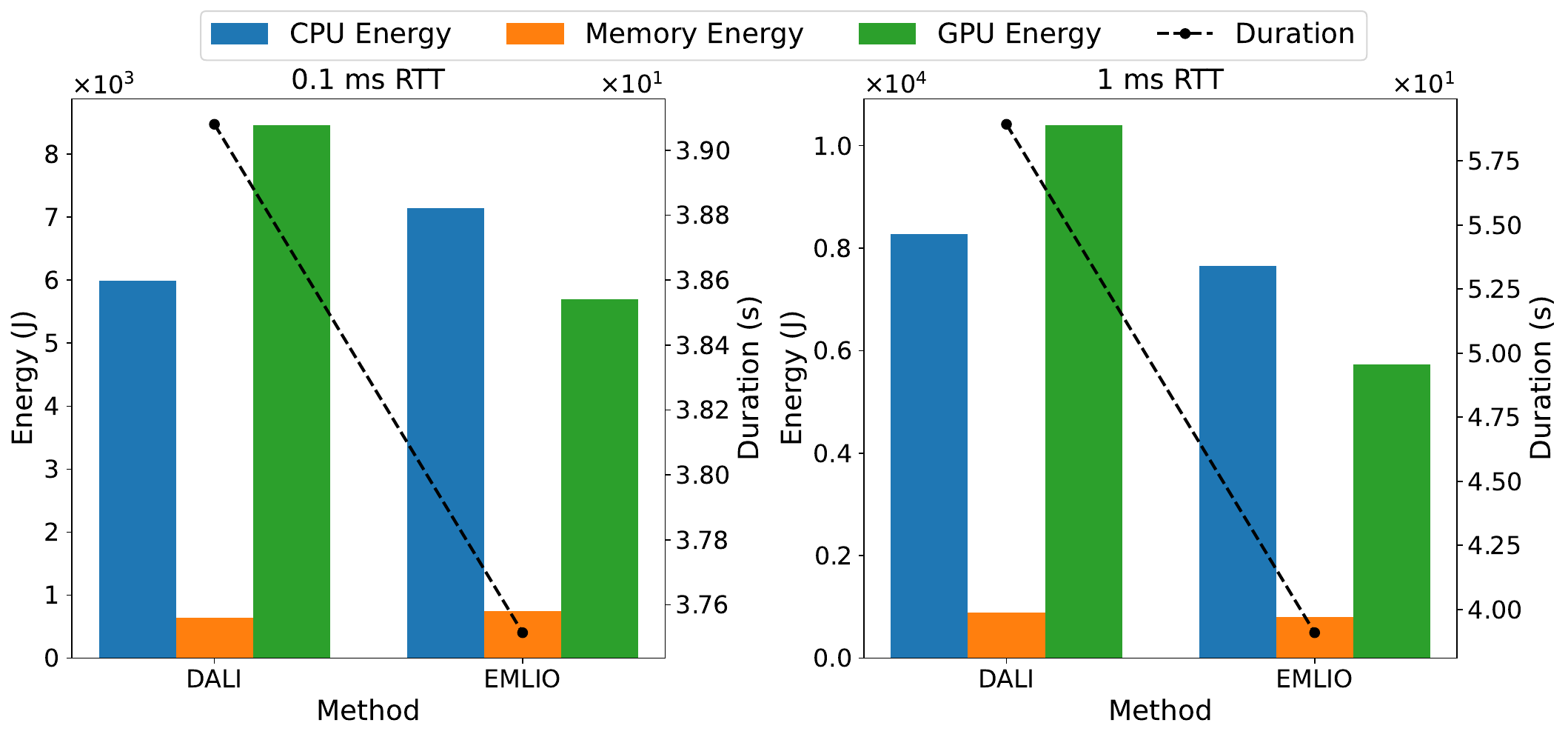}
  \caption{
    Breakdown of per‐epoch energy consumption (CPU, DRAM, GPU) and runtime (line) 
    for NVIDIA DALI, and EMLIO across two network RTT regimes:
    LAN at 0.1 ms, LAN at 1 ms for the synthetic dataset.  EMLIO employs concurrency 2. 
  }
  \label{fig:eval-synth_o.1ms_1ms}
\end{figure}

\paragraph{Synthetic 2 MB Records}  
For large (2 MB) synthetic samples (Figure~\ref{fig:eval-synth}), the one‐stream EMLIO Daemon incurs serialization overhead that briefly makes it slower than DALI at 0.1 ms and 1 ms RTT.  By increasing the daemon’s concurrency as shown in Figure ~\ref{fig:eval-synth_o.1ms_1ms} to two parallel batch‐serialize + send threads, EMLIO amortizes the fixed cost and regains a consistent lead: across all RTTs, EMLIO achieves 2–3× higher throughput and 3–5× lower energy.

These results confirm that:
\begin{itemize}

  \item For \emph{remote} traffic (1–30 ms), EMLIO’s pre‐batching, multi‐stream TCP/ZMQ, and parallel serialization fully hide RTT, yielding essentially RTT‑agnostic performance and energy.
  \item Concurrency in the EMLIO Daemon is key for large records: multiple serialize + send threads overlap compute and I/O to amortize per‐batch setup.
\end{itemize}
Thus, EMLIO provides robust, distance‐invariant data loading, unlike other loaders whose performance and energy deteriorate sharply as network latency increases.

\begin{figure}
  \centering
  \includegraphics[width=\columnwidth]{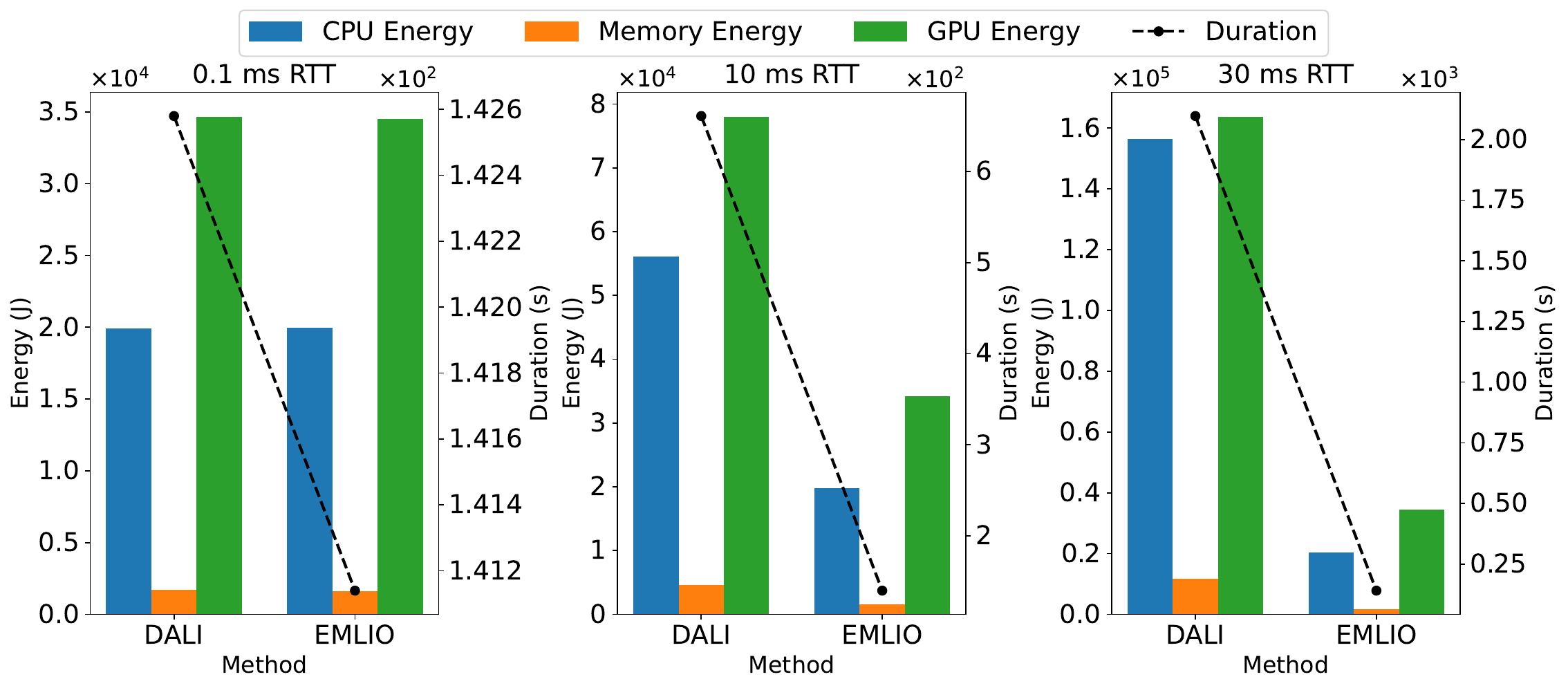}
  \caption{
    Breakdown of per‐epoch energy consumption (CPU, DRAM, GPU) and runtime (line) 
    for NVIDIA DALI, and EMLIO across three network RTT regimes for the VGG-19 model with the Imagenet dataset. 
  }
  \label{fig:eval-imgnet-vgg}
\end{figure}

\paragraph{ImageNet 10 GB Subset on VGG‑19}
We had previously evaluated EMLIO’s performance using the ResNet‑50 model; here we repeat the experiment with VGG‑19 ~\citep{vgg19} to verify that the I/O efficiency gains carry over to a different vision backbone.
Figure~\ref{fig:eval-imgnet-vgg} plots per‑epoch energy (bars: CPU, DRAM, GPU) and runtime (dashed line) for VGG‑19 training on a 10 GB ImageNet subset under three network regimes: LAN (0.1 ms RTT), LAN (10 ms RTT), and LAN (30 ms RTT). Key observations:

\begin{itemize}[leftmargin=*]
\item {LAN (0.1 ms):}
DALI completes in 142.6 s while EMLIO finishes in 141.1 s ($\approx$ 1 \% faster).
Under low latency, both methods have similar energy:
CPU 19.9 kJ, DRAM  1.7 kJ, GPU  34.6 kJ for DALI;
CPU 20.0 kJ, DRAM  1.6 kJ, GPU  34.5 kJ for EMLIO.

\item {LAN (10 ms):}
DALI slows to 660.9 s (4.6×), whereas EMLIO remains at 140.0 s (1.0×).
DALI’s energy rises sharply (CPU  56.1 kJ, DRAM  4.7 kJ, GPU  78.0 kJ),
but EMLIO still uses only  19.8 kJ CPU, 1.6 kJ DRAM, 34.2 kJ GPU.

\item {LAN (30 ms):}
DALI degrades further to 2096.8 s ( 15×), while EMLIO stays at 140.5 s ( 1.0×).
DALI’s I/O energy explodes (CPU  156.3 kJ, DRAM  11.8 kJ, GPU  163.6 kJ),
contrasted with EMLIO’s near‑constant footprint (CPU  20.3 kJ, DRAM  1.6 kJ, GPU  34.4 kJ).
\end{itemize}

Taken together, this confirms that EMLIO’s pipelined, multi‑stream push fully hides network latency and minimizes I/O‑related energy not only for ResNet‑50 but also for VGG‑19, demonstrating that its I/O efficiency advantage generalizes across different vision‑based models.

\subsection{Scenario 2: Data Distributed}

\begin{figure}
  \centering
  \includegraphics[width=\columnwidth]{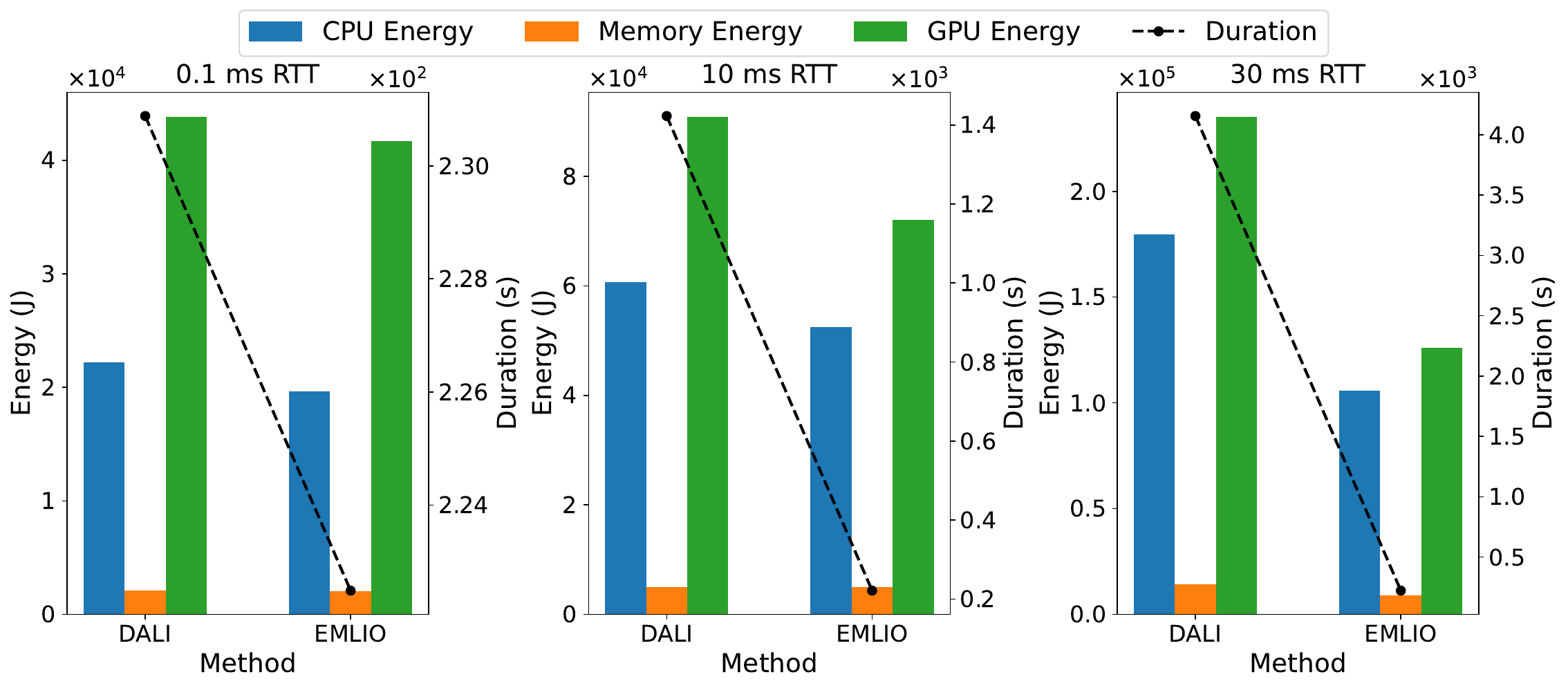}
  \caption{
    Per‑epoch energy consumption (CPU in blue, Memory in orange, GPU in green) and runtime (dashed line) 
    for DALI vs.\ EMLIO+ DALI in the sharded‑dataset scenario (local half + remote half via NFSv4),
    under three emulated RTTs: 0.1 ms, 10 ms, and 30 ms.  
    EMLIO’s batched, multi‑stream transfers hide remote‑read latency, delivering significantly lower time and I/O energy compared to DALI.
  }
  \label{fig:scenario2-results}
\end{figure}

In the sharded–local + remote scenario, the entire training dataset is pre‑sharded and evenly distributed across all compute nodes: each node stores one shard locally but still processes the full dataset by streaming the remaining shards over the network, preserving stochastic gradient descent convergence while balancing I/O load. Figure~\ref{fig:scenario2-results} shows the per‑epoch energy breakdown (CPU, DRAM, GPU) and runtime for DALI vs.\ EMLIO under network RTTs of 0.1ms, 10ms, and 30ms.

\paragraph{0.1 ms RTT}  
In this sharded–local + remote setup, we used the Imagenet dataset and each node reads 50\% of samples from its local shard and 50\% over NFS. With DALI, an epoch takes 230.9 s, consuming \num{2.22e4} J (CPU), \num{2.08e3} J (DRAM), and \num{4.38e4} J (GPU). EMLIO’s pre‑batching keeps DRAM energy nearly unchanged (\num{2.03e3} J, – 2.5 \%), yet cuts CPU energy by 11.5 \% to \num{1.97e4} J and GPU energy by 4.7 \% to \num{4.17e4} J, while shortening runtime by 3.7\% to 222.5 s.

\paragraph{10 ms RTT}  
Under 10 ms latency, DALI’s epoch runtime inflates to 1422.5 s (×6.2), with CPU energy rising to \num{6.07e4} J, DRAM to \num{5.03e3} J, and GPU to \num{9.08e4} J. In contrast, EMLIO completes in 221.6 s—6.4× faster—reducing CPU energy by 13.5 \% to \num{5.25e4} J and GPU energy by 20.8\% to \num{7.20e4} J, while DRAM remains steady at \num{4.96e3} J.

\paragraph{30 ms RTT}  
At 30 ms RTT, DALI’s epoch balloons to 4154.7 s, consuming \num{1.80e5} J (CPU), \num{1.42e4} J (DRAM), and \num{2.35e5} J (GPU). EMLIO still finishes in 221.8 s—18.7× faster—slashing CPU energy by 41.1\% to \num{1.06e5} J, GPU by 46.4\% to \num{1.26e5}J, and DRAM to \num{9.01e3}J.

These results confirm that EMLIO’s batched, multi‐stream transfers remove variable I/O latency, but end‐to‐end epoch time and energy still rise modestly with RTT. This increase is not caused by I/O inefficiency—data loading remains constant—but by higher synchronization overhead across higher‐latency network links.

\subsection{Training Progress Over One Epoch}
\label{sec:training-progress}

\begin{figure}[t]
  \centering
  \includegraphics[width=\columnwidth]{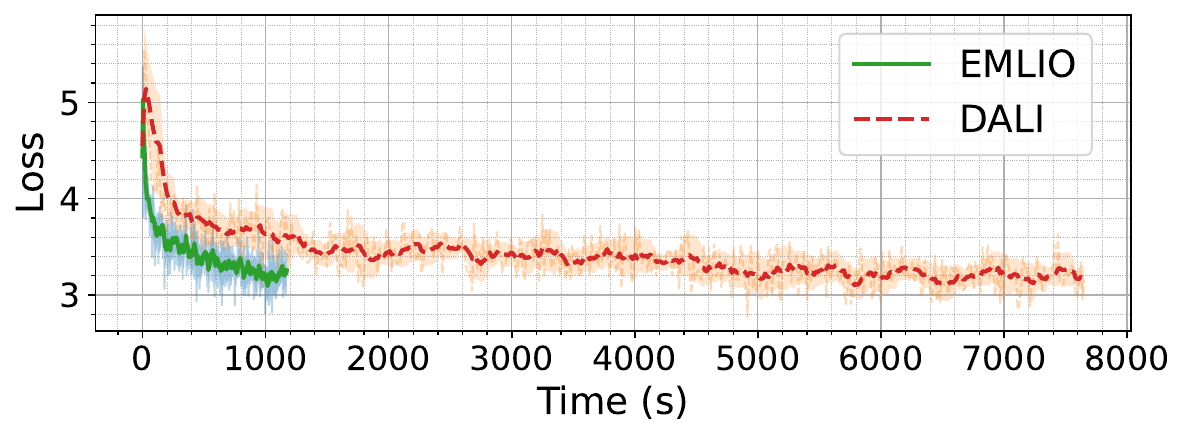}
  \caption{
    Training loss vs.\ wall‑clock time under 10 ms RTT.  
    Solid green: EMLIO, dashed red: DALI.  
    Shaded regions: ±1 std.  
    Vertical grid lines mark 2000 s intervals.  
    EMLIO completes the epoch in 1000 s, vs.7500 s for DALI, and achieves lower loss at every time point.
  }
  \label{fig:training-progress}
\end{figure}

To illustrate how EMLIO’s low‐latency data feed translates into faster convergence in wall‐clock time, we recorded the ResNet‑50 training loss as a function of real time under a 10 ms RTT between compute and COCO dataset storage node.  Figure~\ref{fig:training-progress} plots both the raw per‐iteration loss (thin lines) and a 10‐iteration moving average (thick lines) for NVIDIA DALI (red dashed) and EMLIO (green solid).  Shaded bands show ±1 standard deviation over three runs.

Shortly after start (0–200 s), EMLIO’s loss drops from 5.0 to 3.8, whereas DALI only falls to ~4.0 in the same interval.  EMLIO reaches a final moving‐average loss of 3.2 at around 1000s; DALI, still in mid‐epoch at that point, only attains 3.4 and completes the epoch at 7500s with a loss of 3.3.  This demonstrates that EMLIO’s pipelined, batched data delivery not only reduces epoch duration by over 7× but also accelerates convergence when measured in actual training time.

\section{Conclusion}

We have presented EMLIO, a fully pipelined I/O framework for distributed deep learning that minimizes both data‐loading latency and energy use. By moving pre‑batching and serialization into a storage‐side daemon, streaming fixed‑size TFRecord shards over parallel TCP/ZeroMQ channels, and hooking directly into NVIDIA DALI on the compute nodes, EMLIO delivers near‑constant throughput and energy from 0.1 ms to 30 ms RTT. Our synchronized, fine‑grained energy monitor—built on Linux \texttt{perf stat}, NVML, and NTP‐aligned timestamps—reveals that EMLIO cuts I/O energy by up to 8× and speeds up epochs by up to 13× over PyTorch DataLoader and NVIDIA DALI in both centralized and sharded setups (ImageNet, COCO, and synthetic 2 MB samples).

Future work will focus on three areas: co‑scheduling data loading with DDP gradient synchronization for cross‑layer energy optimization; evaluating heterogeneous transports—such as RDMA and NVMe‑over‑Fabric—to further reduce I/O latency and energy in HPC and cloud environments; and extending EMLIO beyond TFRecord to support additional formats, including audio, text for LLM training, and multimodal vision+language datasets.

\bibliographystyle{ACM-Reference-Format}
\bibliography{reference}

\end{document}